\documentclass[reprint,
prl,
]{revtex4-2}
\usepackage{graphicx}% Include figure files
\usepackage{dcolumn}% Align table columns on decimal point
\usepackage{bm}% bold math
\usepackage{amsmath}
\usepackage{amssymb}
\usepackage{bm}
\usepackage{xr}%\bm=\mathbf
\usepackage{xcolor}
\usepackage{layout}

\makeatletter
\newcommand*{\addFileDependency}[1]{% argument=file name and extension
\typeout{(#1)}% latexmk will find this if $recorder=0
\@addtofilelist{#1}
\IfFileExists{#1}{}{\typeout{No file #1.}}
}\makeatother

\newcommand*{\myexternaldocument}[1]{%
\externaldocument{#1}%
\addFileDependency{#1.tex}%
\addFileDependency{#1.aux}%
}
\myexternaldocument{Supplementary_Material}

\begin{document}

\preprint{APS/123-QED}

\title{Capital Inequality Induced Business Cycles}% Force line breaks with \\

\author{S\"oren Nagel}
\affiliation{Zuse Institute Berlin, Takustr.7, 
14195 Berlin
Germany\\ Potsdam Institute for Climate Impact Research, PO Box 60 12 03, 14412 Potsdam, Germany
}%


\author{Jobst Heitzig}
\affiliation{
 FutureLab on Game Theory and Networks of Interacting Agents, Complexity Science Department, Potsdam Institute for Climate Impact Research, PO Box 60 12 03, 14412 Potsdam, Germany
}%
\author{Eckehard Sch\"oll}
\affiliation{
Institute for Theoretical Physics, Technische Universit\"at Berlin, Hardenbergstr.\,36, 10623 Berlin, Germany,\\ Potsdam Institute for Climate Impact Research, PO Box 60 12 03, 14412 Potsdam, Germany,\\ Bernstein Center for Computational Neuroscience Berlin, Humboldt Universit\"at, 10115 Berlin, Germany
}%

\date{\today}% It is always \today, today,

\begin{abstract}
In this letter we present a stochastic dynamic model which can explain economic cycles. We  show that the macroscopic description yields a complex dynamical landscape consisting of multiple stable fixed points, each corresponding to a split of the population into a large low and a small high income group. The stochastic fluctuations induce switching between the resulting metastable states, and excitation oscillations just below a deterministic bifurcation. The shocks are caused by the decisions of a few agents who have a disproportionate influence over the macroscopic state of the economy due to the unequal distribution of wealth among the population. The fluctuations have a long-term effect on the growth of economic output and lead to business cycle oscillations exhibiting coherence resonance, where the correlation time is controlled by the population size which is inversely proportional to the noise intensity. 
\end{abstract}

\maketitle

%\tableofcontents

%\section{\label{sec:level1}Introduction}
The complex networks approach is a transdisciplinary paradigm to capture the nonlinear dynamics of  a multitude of natural, technological, or social systems.  %After the financial crisis of 2008 there has been an increased effort to develop novel economic models which are able 
In order to predict and help to understand the effects of economic crises or shocks, and guide policymakers to handle such situations\cite{farmer2012complex, ball2014long, Farmer2009-kg}, %To this purpose, 
the economy should be modeled as a complex socioeconomic system with a plethora of network interactions between agents (households) and market institutions, taking into regard that wealth and power are heterogenously and unequally distributed among the population.

The classical approach in economics is to assume complete rationality and only consider a single representative agent, who then solves a long term optimization problem in order to maximize the long term benefits of increased consumption. The typical use of convex functions results in the existence of a unique fixed point that is then disturbed by external shocks  \cite{farmer2012complex, acemouglu2009modern, Asano}. These models in general lack the dynamical complexity needed to describe the economic reality observed, and completely disregard the highly non-uniform distribution of wealth in typical modern economies~\cite{chancel2022world}.
The long-lasting effects of the 2008 financial crisis have lead to the paradigm shift in economics that business cycles and random fluctuations are interdependent in economic growth theory \cite{IMF_review, ABM_hysteresis}. 

This work addresses the question of how stochastic interactions between individuals can give rise to fluctuations of macroeconomic quantities, and the associated long-term effects on economic growth. 

We start from a modified version of the agent-based model for  business cycles and economic inequality presented in \cite{Asano}, but develop a macroscopic stochastic model using the Langevin equation approach \cite{Gillespie, Niemann2021} and a moment closure for a description of the underlying agent-based model for large but finite population size. This approach allows us to first study the deterministic system for an infinite population and then use the Langevin approach to understand the effect of finite size fluctuations. Such macroscopic models are advantageous with respect to  comparing them to data, since they only deal with average and aggregate quantities, which in reality are much easier to obtain than the refined data necessary to specify the initial conditions of an agent-based model. The method could be applied to other problems, e.g., neural systems~\cite{JI23} or network motifs~\cite{BAO22} as well.

{\em Agent-Based ``Micro'' Model.} We study a stochastic model of $N\gg 1$ households  $i$ (``agents'') in a fully connected network, characterized by two dynamic variables $(K_i, S_i)$. Household capital $K_i \geq 0$ is accumulated by saving a fraction of household income given by its current saving rate $S_i$. Although in principle, $S_i$ could be any real number in $[0,1]$, we assume $S_i$ is one of $M>1$ discrete saving rate levels $s_1<\dots<s_M$. Agent $i$ independently and stochastically updates $S_i$ at random times given by a Poisson process with common jump rate $1/\tau$. At each update, $i$ either {\em explores} or {\em imitates.} With probability $\epsilon$, $i$ switches to any of the saving rate levels uniformly at random (``exploration''). With probability $1-\epsilon$, $i$ will instead copy the saving rate $S_j$ of any agent $j$, drawn with a probability that depends on $j$'s current consumption $C_j$ (``imitation''). We assume that the probability to choose agent $j$ for imitation is governed by a Boltzmann distribution with inverse temperature $\beta$,
\begin{align}
    P(S_i \rightarrow S_j) &= \frac{1}{Z}\exp (\beta C_j), & Z &= \sum_{j=1}^N\exp (\beta C_j),
\end{align}
resulting in a voter model with coevolving transition probabilities \cite{BLUME1993387, pris_dilem, evol_game}. In the low temperature limit $\beta \rightarrow \infty$, this ``softmax policy'' converges to the imitate-the-best (``argmax'') policy used in \cite{Asano}, where agents deterministically adopt the saving rate of the agent with the highest consumption in their neighborhood. Our generalization to a stochastic softmax policy can be interpreted as representing rational decision-making under uncertain measurements of others' consumption, similar to \cite{MCKELVEY19956}, and it is a common assumption in behavioral economics and machine learning. 

Household consumption $C_i = (1 - S_i)I_i$ is that part of income $I_i$ which is not saved. Income depends on gross economic production $Y$, determined by a Cobb--Douglas production function \cite{CobbDouglas}
%\begin{align}
$
    Y = A K^{\epsilon_K}L^{\epsilon_L},
$
%\end{align}
where $K=\sum_{i=1}^N K_i$ is aggregate capital, $L$ is aggregate labor, $A$ is a constant, and $\epsilon_L$ and $\epsilon_K$ are elasticities, here $\epsilon_K = \epsilon_L = 1/2$.
Household $i$ supplies their capital and fixed labor $l_i=L/N$ to the economy for production and is compensated at wage $w=\partial Y / \partial L$  and  capital return $r = \partial Y / \partial K$ , resulting in an income \cite{Note1} of
\begin{align}\label{income}
    I_i &= rK_i + wL/N
    = A\sqrt{L}\big(K_i/\sqrt{K} + \sqrt{K}/N\big)/2.
\end{align}
Investing the saved fraction $S_i$ of $I_i$ into capital growth results in a coupled, nonlinear evolution of capital stocks,
\begin{align}\label{k-dot}
    \dot{K_i} = S_iI_i - \kappa K_i = (rS_i - \kappa) K_i - wS_i L /N,
\end{align}
where $\kappa>0$ is the common capital {\em depreciation rate.}

%\subsection{Macroscopic System}
{\em Macro-Model.} To study the agent-based model's oscillatory behavior in the large system limit $N\to\infty$, we focus on a few aggregate quantities: the vector of {\em occupation numbers} $\bm{n} = (n_1,...,n_M)$ of all saving rate levels, and the capital distribution in each of these levels. This admits an approximation via a chemical Langevin equation  \cite{Gillespie, Niemann2021} combined with a moment closure approach for the capital distributions in each saving rate level. 

The Langevin equation incorporates fluctuations in the transition rates due to the finite size of the system. This contrasts the usual ways fluctuations are introduced into macro-economic growth models based on demand shocks, credit defaults, or technological progress \cite{IMF_review}. The time evolution of the occupation numbers follows an It\^o stochastic differential equation (SDE),
\begin{align}\label{SDE-n}
    d\bm{n} = \sum_{k,l=1}^{M} \alpha_{kl} \bm{\nu}_{kl}\, dt + \sum_{k,l=1}^{M} \sqrt{\alpha_{kl}} \bm{\nu}_{kl}\, dB_{kl}.
\end{align}
$\bm{\nu}_{kl} = \bm{e}_k - \bm{e}_l$ indicates a transition between levels $s_k\rightarrow s_l$, where $\bm{e}_k$ is the $k$-th unit vector in $\mathbb{R}^M$, and $dB_{jl}$ are the increments  of uncorrelated white noise.  Due to imitation and exploration, the transition rate for $k \rightarrow l$ is 
\begin{align}
        \alpha_{kl} %&= (1-\epsilon)\frac{n_k}{\tau} \sum_{\lbrace i: S_i = s_l\rbrace}\left[ \frac{1}{Z}\exp (\beta C_i)\right] + \epsilon\frac{n_k}{\tau M} \\
        &=  (1-\epsilon)\frac{n_k}{\tau Z}n_l\langle \exp(\beta C_i)\rangle_l + \epsilon\frac{n_k}{\tau M}. \label{alpha}
\end{align}
where $\langle X_i \rangle_l=(n_l)^{-1}\sum_{\lbrace i: S_i = s_l\rbrace}X_i$ denotes the population average of agents in saving rate level $l$.

For the moment closure, we consider the $p$-th  moment ($p\ge 1$) of the capital distribution among those households whose saving rates are in level $l$: $m_l^p = \langle K^p_i \rangle_l$.  We cannot directly compute the evolution of the capital moments using Eq.~(\ref{k-dot}), since when a household switches to a different saving rate, it takes its capital stock with it. This leads to correction terms \cite{PhysRevE.91.052801, sup_mat, Kuehn2016} that directly couple Eq.~(\ref{SDE-n}) with the evolution of the capital moments,
{\footnotesize
\begin{align}\label{SDE-non-central-moments}
\begin{split}
    dm_l^p &= \left(p(rs_l -\kappa) m_l^p + pws_l \frac{L}{N}m_l^{p-1} + \sum_{k=1}^M \frac{m_k^p - m_l^p}{n_l}\alpha_{kl}\right)\,dt 
    \\
    &+ \sum_{k=1}^M \frac{m_k^p - m_l^p}{n_l}\sqrt{\alpha_{kl}}\, dB_{kl}.
\end{split}
\end{align}
}
Since $\tau\gg 1 $, this results in a slow-fast system, where the occupation numbers are the slow variables. We  apply a Taylor approximation of the exponential in Eq. ~(\ref{alpha}) and  in order to better capture the maximum consumption in each level, we expand about the mean consumption $\langle C_i(t) \rangle_l$,
\begin{align}
    \label{Taylor_expansion}
    &\langle \exp (\beta C_i)\rangle_l = \exp(\beta \langle C_i\rangle _l) \sum_{p=0}^\infty \frac{\beta^p}{p!}\langle \left(C_i-\langle 
    C_i\rangle_l\right)^p\rangle_l.
\end{align}
This reduces the systematic error from underestimating the maximal consumption, when using finitely many terms,  but introduces a further nonlinearity. The moments of the consumption distributions are easily computed from the moments of the capital distribution using $C_i = (1-S_i)I_i$ and Eq.~(\ref{income}), see \cite{sup_mat}.

We choose to include the third moment of the consumption and capital distributions, since the micro-model displays significant amounts of skewness \cite{Asano}. We hence truncate the moment closure at $p=3$. This truncation affects the dynamics only by decreasing the accuracy of the maximum consumption estimate, while the evolution equations for the capital moments are unaffected.

{\em Results.} In our simulations, we restrict ourselves to equidistant saving rate levels $s_l$ within the interval $[0.05,0.95]$.

Let us first explore the basic phase space structure by ignoring the noise terms. %, thereby using a deterministic approximation \cite{Niemann2021} for the macro-scale ($N\rightarrow \infty$, $L/N = const.$ after rescaling $c = n/N$) that gives easy access to.  
For sufficiently large number of levels $M$ and sufficiently high inverse temperature $\beta$, this deterministic approximation displays a complex dynamical landscape with several fixed points, each corresponding to a different distribution of saving rates (Fig. \ref{bifurcations}), see also~\cite{sup_mat}: we observe a split of the population into two groups as in \cite{Asano}. Most agents sit in a low saving rate level with a very low capital stock, but a small group of agents has very high saving rates and owns most of the capital; this state is prominent for all parameters considered. This gives the few ``high savers'' crucial influence on the overall dynamics. 

\begin{figure}
\includegraphics[width=.48\textwidth]{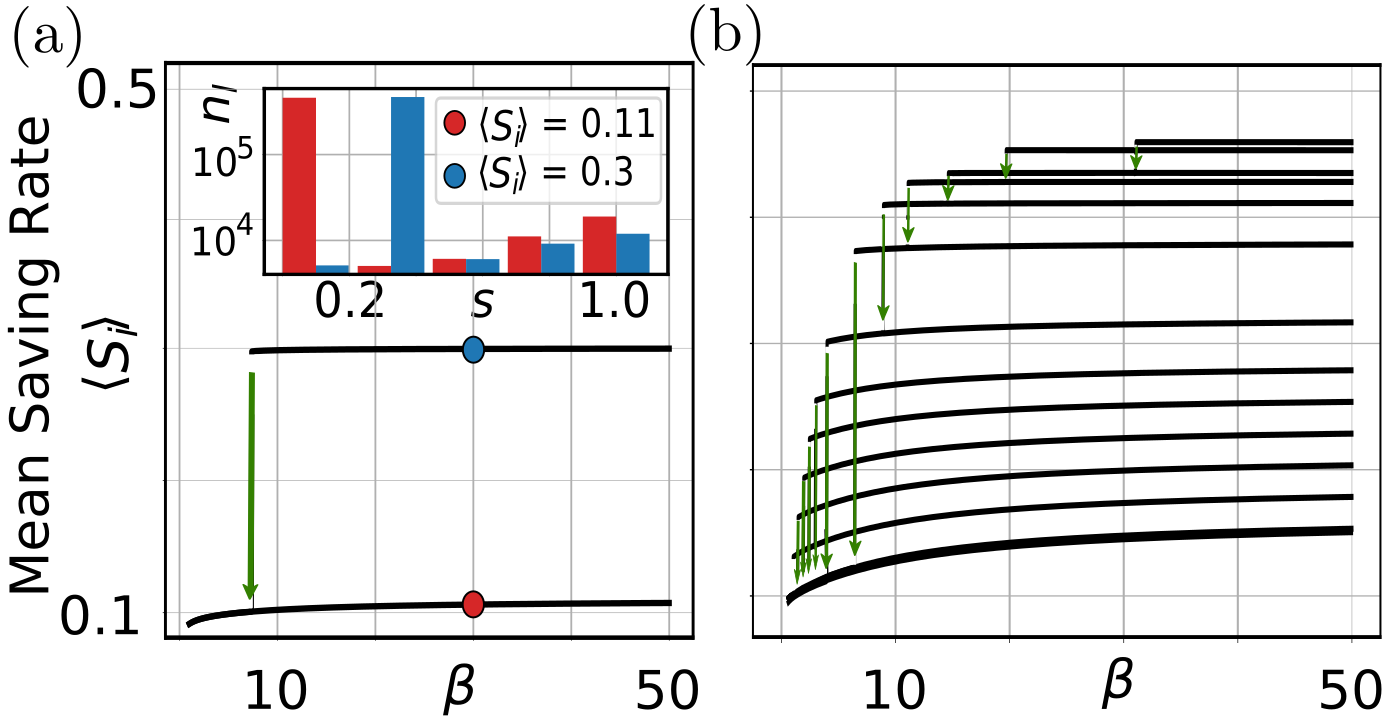}
% Here is how to import EPS art
\caption{\label{bifurcations} Bifurcation diagram of the deterministic approximation for (a) $M=5$ levels: For small $\beta$, there is only one stable state with a very low mean saving rate. For $\beta> 7.35$ another stable state appears with a higher mean saving rate. The inset shows the two saving rate distributions (red/blue) associated with the two states. (b) For $M=30$ there is a cascade of bifurcations. This highlights the possibility of irreversible tipping transitions indicated by green arrows. Parameters: $L=N= 2.5 \cdot 10^5,\,\tau=300,\,\kappa=\epsilon=0.05,\, A = 1$.}
\end{figure}

\begin{figure}
\includegraphics[width=.48\textwidth]{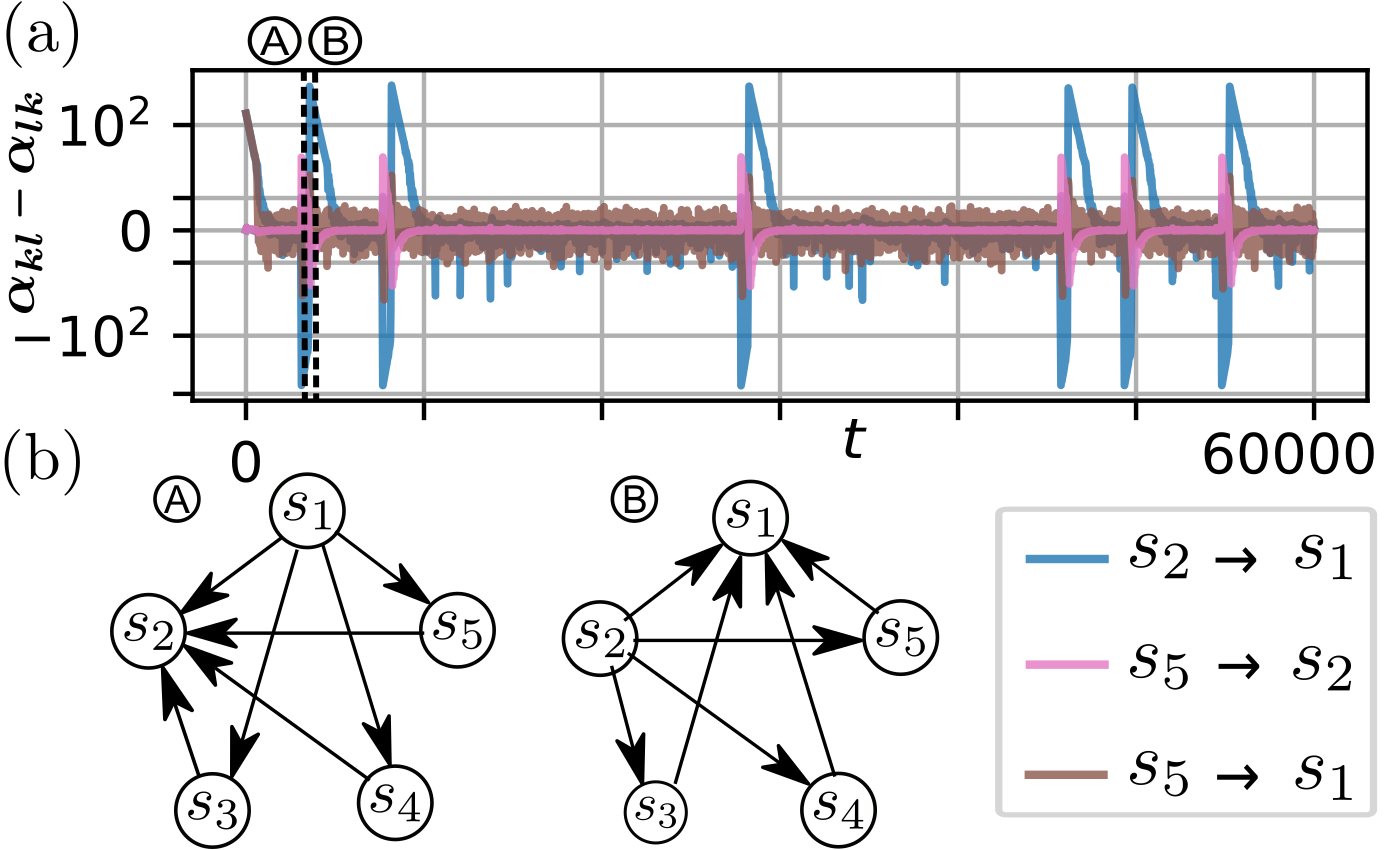}
\caption{\label{transition rates} (a) Net transition rates $\alpha_{kl} - \alpha_{lk}$ between the saving rate levels, just below the bifurcation in Fig \ref{bifurcations}(a). The system displays excitation oscillations, where each oscillation starts with agents switching from $s_1 \rightarrow s_2$ and $s_5\rightarrow s_2$. The arrows in the legend denote positive net transition rates.
(b) Direction of the net flows at times A (excitation) and B (recovery). Although most agents transit from $s_1=0.05$ to $s_2 = 0.27$ during the spike, the path they take can be complicated.  After the majority of agents have changed their saving rate to $s_2$, the flows turn around and almost symmetrically bring the system back to the original distribution of saving rates. Parameters as in Fig.\ref{bifurcations}(a) with $M=5$ and $\beta=5$.}
\end{figure}

With the addition of noise, we switch to a meso-scale for a large but finite number of  households. Here the multi-stability on the macro-scale results in excitation oscillations and switching between the now metastable states.

 In the following we will focus on the case $M=5$. Above the bifurcation in Fig.\ref{bifurcations}(a) the fluctuations lead to switching between the metastable states. Just below the bifurcation, i.e. below the critical inverse temperature $\beta$, the system takes an excursion through the phase space, before returning to the original stable state. This is evidenced by the net transition rates shown in  Fig. \ref{transition rates}.

Therefore, the presence of intrinsic fluctuations induces macroeconomic shocks. This stands in contrast to the deterministic model, which does not produce these shocks.

The split  of the population into two groups of agents with high and low capital stock, respectively, leads to a disparity of influence that drives the transitions between the two metastable states of the system.

Each switching transition is preceded by an abrupt spike in mean consumption in the saving rate level to which  the agents then switch. This is shown in Fig. \ref{few_agents_cause transitions} for $\beta$ above the deterministic bifurcation but otherwise same parameters as in Fig. \ref{transition rates}, and for two saving levels $s_2$ (a,c,e) and $s_5$ (b,d,f).  This spike exponentially increases the transition rate into that level according to Eq.\ref{alpha}. The preceding spike for a switch to higher mean saving rate is depicted in (a) for level $s_2$, which most agents will finally adopt, and in (b) for the highest saving rate level $s_5$. The blow-ups (c), (d) show that the spike in $s_2$ happens on a much faster timescale than the spike in $s_5$ and cannot be attributed to changes in the economic variables, i.e., the market dynamics, since the capital return $r$ (and thus also the wages) is almost constant during the spike. Panel (e) visualizes that during the spike significant amounts of capital are transferred by a small amount of agents (0.1\% of the population \cite{sup_mat}) switching from $s_5$ to $s_2$ and all other capital flows are significantly smaller or reduce the average capital in $s_2$. (f) shows that this capital flow leads to the increase in average consumption, which then draws all the agents into this level. 

This shows the disparity of influence, which is generated from the average capital difference for high and low savers. The decisions of a tiny fraction of influential agents can lead to tipping of the entire macroeconomy.

\begin{figure}
\includegraphics[width=.48\textwidth]{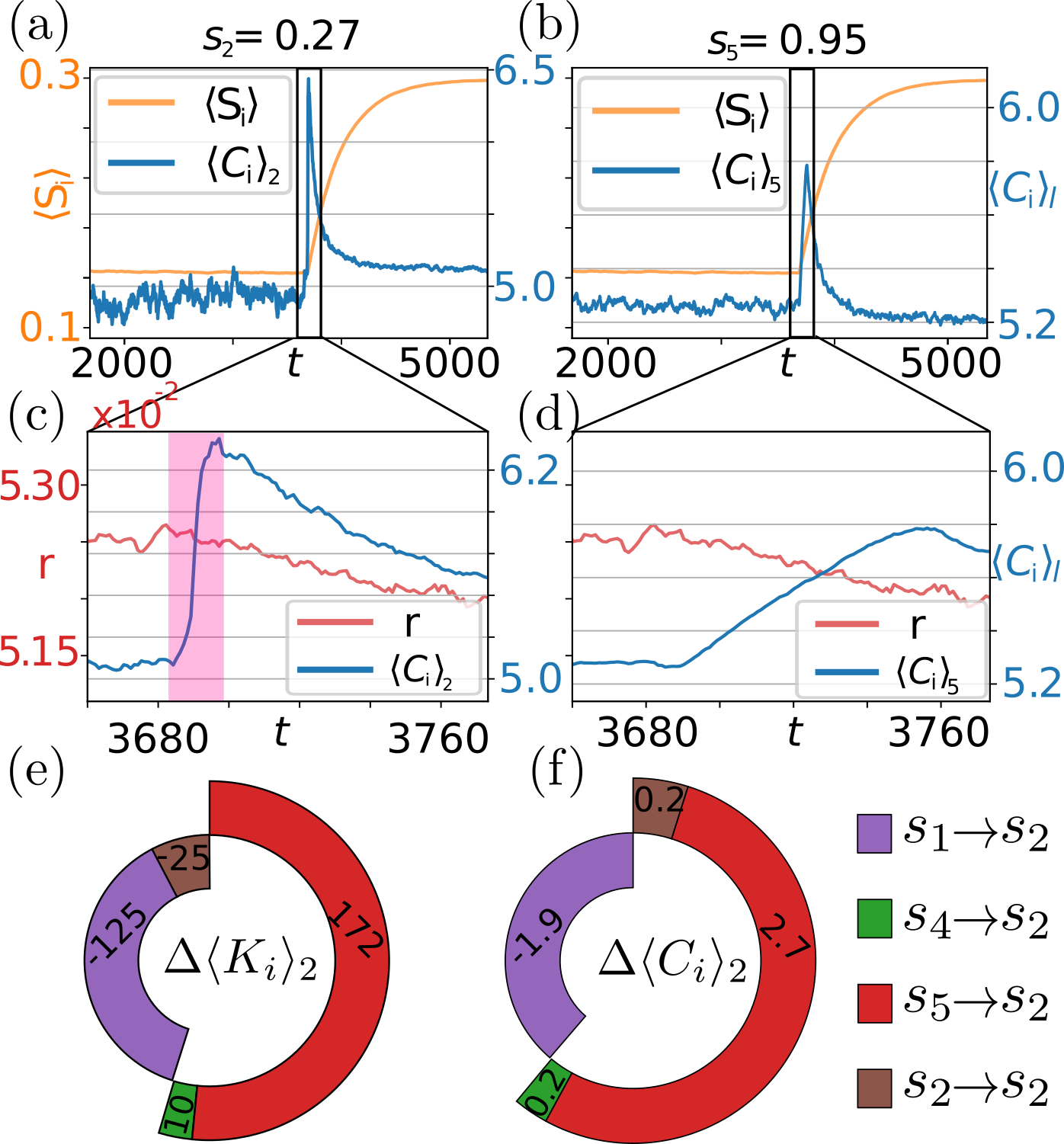}
\caption{\label{few_agents_cause transitions} Mean consumption $\langle C_i(t) \rangle_l$ in the saving rate levels $s_2$ (a,c,e) and $s_5$ (b,d,f), and mean saving rate $\langle S_i(t) \rangle$ . (c) and (d) are blow-ups of (a), (b), respectively, showing also the return $r(t)$. (e) Change of capital $\langle K_i \rangle_2$ and (f) of consumption $\langle C_i \rangle_2$ in  level $s_2$ due to transitions from the other levels. (The transition $s_3 \rightarrow s_2$  is negligible and invisible.) Parameters as in Fig.\ref{bifurcations}(a) with $M=5,\, \beta = 15$.}
\end{figure}

The small increase of "high savers" moving to the lower saving rate is due to the finite size fluctuations of the transition rates, which are then amplified by the economic inequality and the fact that the level $s_2$ has a very low occupation. This combination creates a timescale separation, allowing for the sudden increase in mean capital. 

The consumption spike in the high saving rate level $s_5$ in Fig. \ref{few_agents_cause transitions} happens on the much slower timescale which is associated with the economic variables, mainly the depreciation rate $\kappa$. The change in occupation numbers then follows on the slow timescale.

%%%%%%%%%%%%%%%%%%%%%%%%%%%%%%%%%%%%%%%%%%%%%%%%%%%%%%%%%%%%%%%%%%%%%%%%

%There is a discussion in the economics community about the long-term effects of production fluctuation \cite{IMF_review, ABM_hysteresis}. 

%As outlined in \cite{IMF_review, ABM_hysteresis} there has been a recent paradigm shift towards studying the \emph{hysteresis} behavior in business cycles, i.e. the long term effect of fluctuations on the growth rate of economic production.  This concept has only recently been of interest in economics, although there still is some debate on whether hysteresis is actually  a real world effect.

Our model exhibits hysteresis of business cycles, referring to long-term effects of fluctuations on economic growth, which corroborates a recent paradigm shift in economic growth theory \cite{IMF_review, ABM_hysteresis}. We find  that fluctuations can lead to a long term change in production, if the system is in the multistable regime and population growth is introduced
~\cite{acemouglu2009modern}. For details and empirical literature see \cite{sup_mat}. 

\emph{Coherence resonance} is a common phenomenon in noisy excitable nonlinear systems. It describes the non-monotonic dependence of the coherence of noise-induced oscillations upon noise intensity, i.e., there exists an optimum noise intensity maximizing coherence \cite{Pikovsky}. A common measure of coherence is correlation time \cite{Delay_Feedback_Controls_Noise, zakharova2013coherence,Geffert2014}. Coherence resonance is visible as maximum correlation time~\cite{sup_mat} for non-zero noise intensity.

Since population growth is a major driver of economic growth, it is natural to ask how the oscillatory behavior changes as  the system size increases. After rescaling Eqns.~{(\ref{SDE-n}), (\ref{SDE-non-central-moments})} to densities $\bm{c}=\bm{n}/N$, the population size $N$ directly corresponds to a noise intensity parameter $\Gamma = 1/\sqrt{N}$ if $L/N=const.$

In Fig. \ref{correlation time} the correlation time of the economic production is plotted vs. noise intensities for several inverse temperatures $\beta$. For $\beta=5$, i.e., below the bifurcation in Fig.\ref{bifurcations}(a) the correlation time exhibits a very flat region of increased coherence. Above the bifurcation ($\beta=8$) the correlation time increases drastically.
For $\beta=50$ we have a clear peak of optimum coherence at $\Gamma \approx 1.7 
\times 10^{-3}$, and a very broad second maximum upon further increase of noise intensity. Note that no deterministic bifurcation is involved in the dramatic change of the behavior of the correlation time, when going to larger inverse temperatures. However, the time spent near each metastable state changes drastically. (see Fig. 2(b),(c) in \cite{sup_mat}).

 \begin{figure}
\includegraphics[width=.48\textwidth]{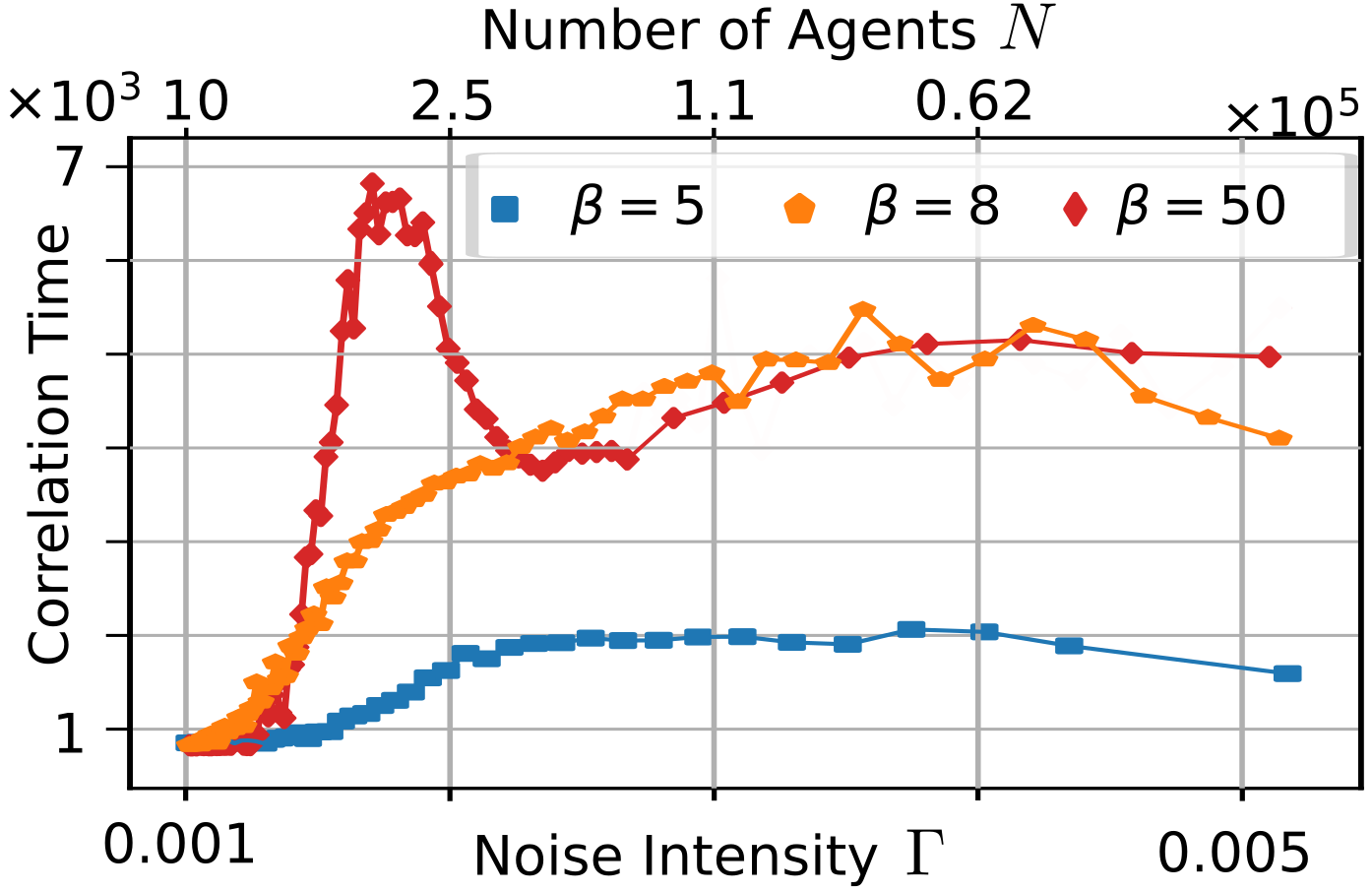}
\caption{\label{correlation time} Correlation time for the economic production $Y$ vs noise intensity $\Gamma = 1/\sqrt{N}$  for different values of $\beta$ below the bifurcation ($\beta=5$) and above the bifurcation  ($\beta=8$, $\beta=50$). For $\beta=50$ the peak at optimum noise intensity $\Gamma \approx 1.7\times 10^{-3}$ or $N \approx 3.5\times 10^{5}$ indicates coherence resonance. Parameters as in Fig.\ref{bifurcations} with $L=N$.}
\end{figure}
This illustrates that the precision, with which the agents imitate the behavior of the agents with the highest consumption, can have a strong effect on the coherence of the business cycle oscillations. 
The presence of coherence resonance indicates that fluctuations in the system can dramatically affect the business cycles and make them more coherent at a certain noise intensity defined by the size of the population.  In more realistic models stochastic fluctuations can arise from other sources as well, and although %care should be taken since 
we are dealing with multiplicative noise, it seems plausible that other sources of noise might lead to oscillatory behavior that has similar characteristics, because the underlying phase space structure strongly influences the response to fluctuations.

In conclusion, we have developed a macroscopic model which captures the characteristic dynamical features of the agent-based model proposed in \cite{Asano}, and beyond that includes specific effects of stochastic fluctuations like coherence resonance.
The decision-making process of the households creates a high degree of multistability in the infinite population limit, particularly when many saving rate levels are allowed. The multistable states result in a split of the population into a small group with high saving rate and high capital, and a large group of low savers with low capital stock, where the smaller group can exert great influence on the entire economy.

The effect of finite-size fluctuations arising in the case of a large but finite population size leads to the possibility of excitation oscillations and stochastic switching between metastable states, which correspond to a synchronized change of saving strategy of a majority of agents in the population. Compared to \cite{Asano}, the increased population size results in business cycles that are much more abrupt and more akin to rare isolated events than sustained oscillations.

In going beyond the agent-based model \cite{Asano} we are able to deal with a substantially larger population and show that the capital inequality leads to timescale separation, which can cause rapid changes in macroscopic variables. We also find that only about $0.1\%$ of the population~\cite{sup_mat} are responsible for triggering a recession period.

With the introduction of economic growth through a growing population, we show that the fluctuations can lead to long-lasting recessions in economic production, which is commonly discussed as hysteresis in the economics community. Hysteresis of business cycles has typically been linked to fluctuations in financing, debt and monetary policy, and only in a few cases with heterogeneous agents~\cite{sup_mat}. Our model is considerably simpler, and explains the metastability as well as the fluctuations solely from the decision-making process of heterogenous households, in contrast to external sources of noise. To the best of our knowledge, such long-term effects on growth solely resulting from the collective saving behavior of households have not been noted before.

In our model, we find coherence resonance and a qualitative change in the correlation time when the system switches from excitation oscillation to the stochastic switching regime for larger $\beta$, which may elucidate also the effects of other sources of stochastic fluctuations.

\begin{acknowledgments}
This work was supported by DFG (German Research Foundation) - Projects No. 429685422 and 440145547 and under Germany’s Excellence Strategy through grant EXC-2046 The Berlin Mathematics Research Center MATH+ (project no. 390685689).
\end{acknowledgments}
\providecommand{\noopsort}[1]{}\providecommand{\singleletter}[1]{#1}%
%

%%%%%%%%%% Merge with supplemental materials %%%%%%%%%%
\pagebreak
\widetext
\begin{center}
\textbf{\large Supplemental Materials: Capital Inequality Induced Business Cycles}
\end{center}
%%%%%%%%%% Merge with supplemental materials %%%%%%%%%%
%%%%%%%%%% Prefix a "S" to all equations, figures, tables and reset the counter %%%%%%%%%%
\setcounter{equation}{0}
\setcounter{figure}{0}
\setcounter{table}{0}
\setcounter{page}{1}
\makeatletter
\renewcommand{\theequation}{S\arabic{equation}}
\renewcommand{\thefigure}{S\arabic{figure}}
\renewcommand{\bibnumfmt}[1]{[S#1]}
\renewcommand{\citenumfont}[1]{S#1}
%%%%%%%%%% Prefix a "S" to all equations, figures, tables and reset the counter %%%%%%%%%%
\title{Supplementary Material on:\\ Capital Inequality Induced Business Cycles}% Force line breaks with \\

\author{S\"oren Nagel}
\affiliation{Zuse Institute Berlin, Takustr.7, 
14195 Berlin
Germany\\ Potsdam Institute for Climate Impact Research, PO Box 60 12 03, 14412 Potsdam, Germany
}%

\author{Jobst Heitzig}
\affiliation{
 FutureLab on Game Theory and Networks of Interacting Agents, Complexity Science Department, Potsdam Institute for Climate Impact Research, PO Box 60 12 03, 14412 Potsdam, Germany
}%
\author{Eckehard Sch\"oll}
\affiliation{
Institute for Theoretical Physics, Technische Universit\"at Berlin, Hardenbergstr.\,36, 10623 Berlin, Germany,\\ Potsdam Institute for Climate Impact Research, PO Box 60 12 03, 14412 Potsdam, Germany,\\ Bernstein Center for Computational Neuroscience Berlin, Humboldt Universit\"at, 10115 Berlin, Germany
}%

\date{\today}% It is always \today, today,

\maketitle

\section{Derivation of the macroscopic system}
The chemical Langevin equation is a well known approximation for agent-based micro models with fully connected networks \cite{Niemann2021_SM} and with discrete agent states $S_i$ of agent $i$, which are the available saving rates levels $s_1, ..., s_M$ in our case. Having defined a set of allowed saving rate levels, we can average the transition probabilities
\begin{align}
    P(S_i\rightarrow S_j) = \frac{1}{Z}\exp(\beta C_j) \qquad \text{with}\, Z=\sum_{j'=1}^N\exp(\beta C_{j'}),
\end{align}
over the resulting subpopulations of agents with identical saving rate, to find the transition rates $\hat{\alpha}_{kl}$ between the saving levels $k$ and $l$ 
\begin{align} \label{individual transition prob}
   \hat{\alpha}_{kl}&=\frac{1}{\tau}\sum_{\substack{i:S_i=s_k\\j: S_j=s_l}}P(S_i\rightarrow S_j)  =\frac{1}{\tau}\sum_{i:S_i=s_k} n_l \langle P(S_i\rightarrow S_j) \rangle_l = n_k\frac{n_l}{\tau Z}\langle \exp (\beta C_j)\rangle_l,
\end{align}
for the imitation behavior. Here $Z = \sum_{l' = 1}^M n_{l'} \langle \exp (\beta C_j)\rangle_{l'}$ is the normalizing factor, ensuring $ \sum_{k',l' = 1}^M \hat{\alpha}_{k'l'} = N/\tau$, which is the rate of the combined Poisson processes from each agent. Adding the simple transition rates for the exploration behavior, we get the complete transition rates

\begin{align}
        \alpha_{kl} %&= (1-\epsilon)\frac{n_k}{\tau} \sum_{\lbrace i: S_i = s_l\rbrace}\left[ \frac{1}{Z}\exp (\beta C_i)\right] + \epsilon\frac{n_k}{\tau M} \\
        &=  (1-\epsilon)\frac{n_k}{\tau Z}n_l\langle \exp(\beta C_i)\rangle_l + \epsilon\frac{n_k}{\tau M}. \label{alpha_sm}
\end{align}
These are needed for the stochastic differential equation (SDE) of the occupation numbers
\begin{align}\label{SDE-n-SM}
    d\bm{n} = \sum_{k,l=1}^{M} \alpha_{kl} \bm{\nu}_{kl}\, dt + \sum_{k,l=1}^{M} \sqrt{\alpha_{kl}} \bm{\nu}_{kl}\, dB_{kl}.
\end{align}

To account for the fact that agents switching between saving levels take their capital with them, we derive an SDE for the moments $\lbrace m_l^p\rbrace_{p=1}^\infty$ of the capital distributions of agents in each level $l$, which can be performed in a similar way as in \cite{correction_terms_paper}. Indeed, this can be done for any quantity $X_i(t)$ that is associated with the agents $i$ and satisfies a differential equation $\dot{X}_i = F_l(X_1, ... X_N)$ if agent $i$ is in level $l$ and jumps.  Let $J_l(t)\subseteq \lbrace1,...,N\rbrace$ denote the index set of agents with savings rate $s_l$ at time $t$ and  $f_l(t) = \langle F_l(X_i)\rangle_l = \sum_{i\in J_l(t)}F_l(X_i)$ the averaged evolution equation. Then we can write the time evolution of the averaged quantity $x_l(t) := \frac{1}{n_l(t)}\sum_{i \in J_l(t)} X_i(t)$ as
\begin{align}
    d\,x_l(t) &\approx x_l(t+t') - x_l(t) = \frac{\sum_{i \in J_l(t+t')} X_i(t+t') - n_l(t+t')x_l(t)}{n_l(t+t')}.
\end{align}
Now we have the sum over the values $X_i(t +t')$ which are associated with the agents in level $l$ at time $t+t'$. To account for the agents changing saving rate in the time interval, we can split this into the three useful sets: $J_l(t)$ contains those agents that already were in level $l$ at time $t$, $J_l(t+t')\setminus J_l(t)$ contains those agents that arrived in that time interval, and $J_l(t)\setminus J_l(t+t')$ contains those agents that left in that time interval. Then
\begin{align}
    \sum_{i \in J_l(t+t')} X_i(t+t') &=\quad\, \sum_{i \in J_l(t)}\qquad \left[ X_i(t) + dt \dot{X_i}(t)\right]\nonumber\\
    &+ \sum_{i\in J_l(t+t')\setminus J_l(t)} \left[ X_i(t) + dt \dot{X_i}(t)\right]\nonumber \\
    &- \sum_{i\in J_l(t)\setminus J_l(t+t')} \left[ X_i(t) + dt \dot{X_i}(t)\right]\\
    = n_l(t) \left[ x_l + dt\, f_l\right] &+\sum_{k\neq l}\left( dt\,\alpha_{kl}+\sqrt{\alpha_{kl}}\,dB_{kl}\right)  \left[ x_k + dt\, f_k\right]\nonumber\\ &-
    \sum_{k\neq l} \left( dt\,\alpha_{lk}+\sqrt{\alpha_{lk}}\,dB_{lk}\right)\left[ x_l + dt\, f_l\right],
\end{align}
where we have used the transition rates, $\alpha_{kl}$, to approximate the number of agents switching in a small  time interval $[t, t+t']$. This is valid, since the underlying stochastic process assumes that agents are uniformly picked at random to update their saving rate. This means that for any small enough time interval the agents switching between levels have the same distribution as the levels themselves. Now we omit terms of order $dt^2$ and $dt\,dB_{kl}$, which is valid since we are interested in the limit $t'\rightarrow 0$.
We get
\begin{align}
    x_l(t+t') - x_l(t) &= \frac{1}{n_l(t+t')}\bigg(-x_l(t)\left[n_l(t+t')-n_l(t)\right] + n_l(t)f_l(t)dt \\ &\left. +\sum_{k\neq l}\left( dt\,\alpha_{kl}+\sqrt{\alpha_{kl}}\,dB_{kl}\right)  x_k-
    \sum_{k\neq l}  dt\,\alpha_{lk}+\sqrt{\alpha_{lk}}\,dB_{lk}x_l)\right)
\end{align}
and expanding $n_l(t+t')-n_l(t)$ in the numerator by Eq.~(\ref{SDE-n-SM}) and taking $t'\rightarrow 0$ we obtain
\begin{align}
    d\,x_l = f_l(t) dt + \sum_{k=1}^M \frac{x_k(t)-x_l(t)}{n_l(t)}\left(\alpha_{kl}dt + \sqrt{\alpha_{kl}}dB_{kl}\right)
\end{align}
Notably, agents leaving a level do not have an impact on that level's distribution. These additional terms couple the stochasticity of the agent-based model to the market dynamics, which gives rise to the excitation oscillations.

Using the population averages for the capital stock, we can find the evolution equations for the capital moments $m_l^p = \langle K_l^p\rangle$. From Eq.~(\ref{k-dot}) in the main text, we get:
\begin{align}\label{economic-dynamics}
    f_l^p &= \langle\frac{d}{dt}K_i^p \rangle_l = \langle p K_i^{(p-1)} \dot{K_i} \rangle \\
    &= p (rs_l -\kappa)\langle K_i^p\rangle + pws_l L/N \langle K_i^{(p-1)} \rangle
\end{align}
Combining this result with the additional terms due to switching and the Taylor approximation Eq.~(\ref{Taylor_expansion}) from the main text gives the final closed system of SDEs, where the moments of the consumption distribution, which are needed for the Taylor approximation, are easily computed from using $C_i = (1-S_i)I_i$ and Eq.~(\ref{income}) in the main text
\begin{align}
    \langle C_i^p\rangle_l &= (1-s_l)^p\sum_{\rho=0}^p \binom{p}{\rho} r^\rho m_l^\rho (wL/N)^{p-\rho}
\end{align}
The non-linearity in the wage $w$ and capital return $r$ only depend on aggregate capital $K$, 
\begin{equation}
    K = \sum_{i =1}^N K_i = \sum_{l=1}^L\sum_{\lbrace i\vert S_i=s_l\rbrace} K_i = \sum_{l =1}^{L} n_l \langle K_i \rangle_l, 
\end{equation}
which only depends on macroscopic variables, so there is no need to omit terms when considering only a finite number of moments.
\section{Time Scale Separation}
\begin{figure}
    \centering
    \includegraphics{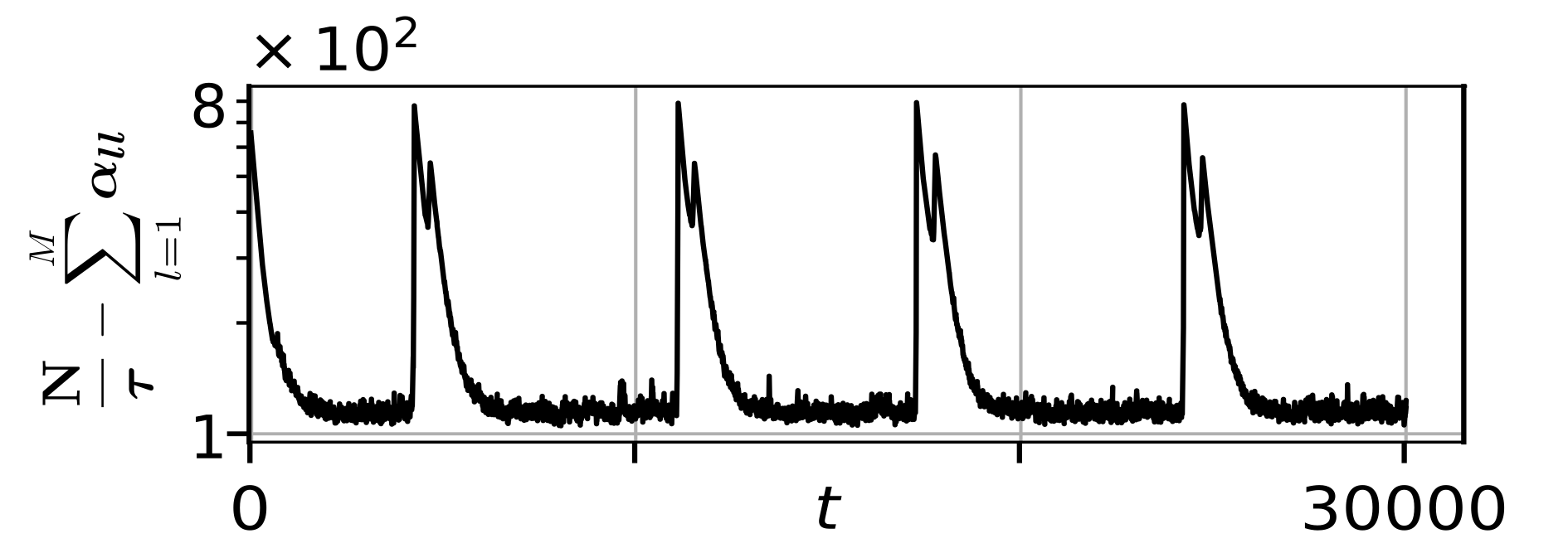}
    \caption{Updates of the saving rate that actually result in a changed saving rate of each agent.  The measurable activity of agents in between excitations is very low and households in general stick to their saving rate. When an excitations occurs, a majority of agents in the system synchronously start imitating agents with a different saving rate. Parameters as in Fig.\ref{transition rates} in the main text.}
    \label{net_flows}
\end{figure}

We already discussed that the large update time $\tau\gg 1$ leads to a slow-fast system. However, the coupling between decision process and market dynamics creates multiple time scales. The fastest time scale arises from the combination of capital inequality and agents changing their saving rate, which is suppressed near the fixed points, because most agents do not actually change their saving rates when imitating. 
The rate at which agents change their saving rate is not $N/\tau$ which is the rate at which decisions are made (since we have $N$ Poisson processes with rate $1/\tau$) but this rate is reduced by the momentary rate at which they choose to imitate another agent with the same saving rate
 \begin{align}
     \frac{N}{\tau} - \sum_{l=1}^M\alpha_{ll}
 \end{align}
 which is illustrated in Fig.\ref{net_flows}.
 
 While the system is near a metastable state, most agents choose to imitate an agent with the same saving rate, but when an excitation occurs suddenly most households choose to change their saving rate instead of remaining in their current state (see Fig. \ref{transition rates} in the main text). So the 
 excitations can be seen as an expression of uncertainty in the population.

The next fastest timescale is the normal economic dynamics given by Eq.~(\ref{economic-dynamics}) in the main text, followed by the dynamics of the occupation numbers, as illustrated in Fig.\ref{few_agents_cause transitions} of the main text. The slowest timescale is the switching between metastable states and the excitation oscillations. In the main text we briefly mentioned that a lot of time can pass between economics shocks. In Fig. \ref{hysteresis}(b, c) we show the histograms for the resting times near each metastable state, for the same system considered in Fig.\ref{bifurcations} in the main text. For $\beta=50$ we see resting times above $250\tau$ even for a relatively small number 287 of observations. Also, the resting time distributions depend on the metastable state from which the process escapes, which illustrates the importance of including multiplicative noise that controls the noise intensity near the fixed points.

Although there are no bifurcations between $\beta=15$ and $\beta=50$, we observe a drastic change of the resting time distribution. As $\beta$ increases, the expected resting time for both states increases and for higher $\beta$ the system spends more time in the state with higher average saving rate, which is generally desirable, since it generates a higher level of economic output and also implies less economic inequality.

\section{Hysteresis of business cycles}
\begin{figure}
    \centering
    \includegraphics[width=\textwidth]{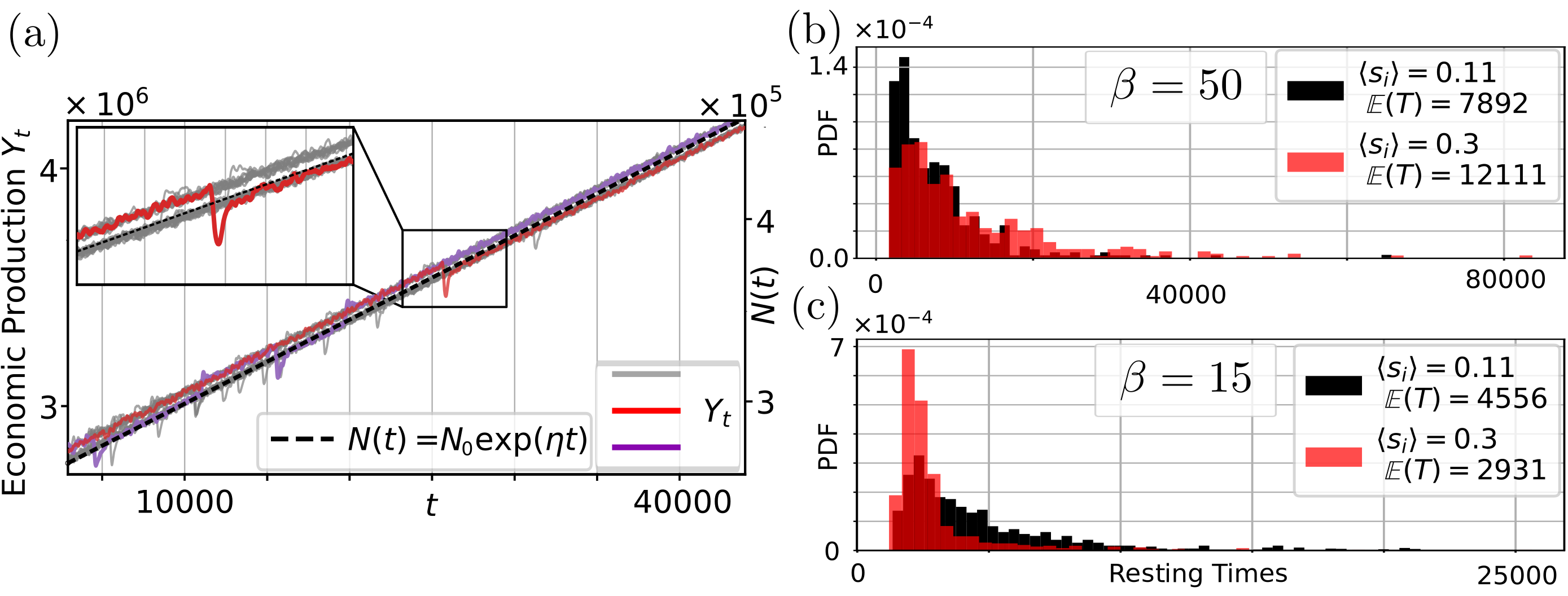}
    \caption{\label{hysteresis} (a) Economic production $Y_t$ vs time. Shocks can have long-lasting effects on the economic growth of an expanding population. For a large enough population, the noise intensity is not sufficient to induce further transitions and the system will remain in one of the stable states. The red and purple realizations show the case where the economy settles in one of the two states, grey:  30 other realizations. Moving averages over a window size of 60 are plotted. The inset shows a blow-up. Parameters as in Fig.\ref{bifurcations}(a) in the main text with $\beta=35,\, L(t) = N(t)$ growing with $\eta=1.6\cdot10^{-3}$ from $N_0= 2.5\cdot10^5$. (b, c) Histograms for time the system remains near the metastable states for (b) $\beta=50$ and (c) $\beta=15$ with 287 and 750 observed resting times respectively. For $\beta=50$ the high saving rate state is occupied more often than for $\beta=15$. The resting times in each state can be very long without any indication of an upcoming shock. Parameters as in Fig.\ref{correlation time} in the main text.}
\end{figure}
Hysteresis of business cycles in economics refers to the long term effects of economic shocks on economic growth \cite{IMF_review_SM}. We consider an exponentially growing population while keeping $L/N = l_0= const.$, which is a classical approach to introduce economic growth  \cite{acemouglu2009modern_SM}. In the multistable regime (large $\beta$) we immediately obtain different degrees of growth for the different states, since they correspond to different values of aggregate capital and thus production.

In Fig. \ref{hysteresis}(a) we see that for a finite population, the fluctuations in the system are strong enough to excite the system to switch between the different states, and that there are realizations of the stochastic process where switching is a rare economic event. Since we assume that the agents' decision when to switch their saving rate is uncorrelated, the noise intensity is proportional to $\sqrt{1/N}$ (Eqs.~(\ref{SDE-n}), (\ref{SDE-non-central-moments}) in the main text). Hence, as the population grows, the fluctuations decrease, and transitions between the multistable states become less likely.
So essentially the economy will settle into one of these states and the future growth rate will be fixed (Fig. \ref{hysteresis}).

\section{The Case with more Available Levels}
\begin{figure}
    \centering
    \includegraphics[width=0.7\textwidth]{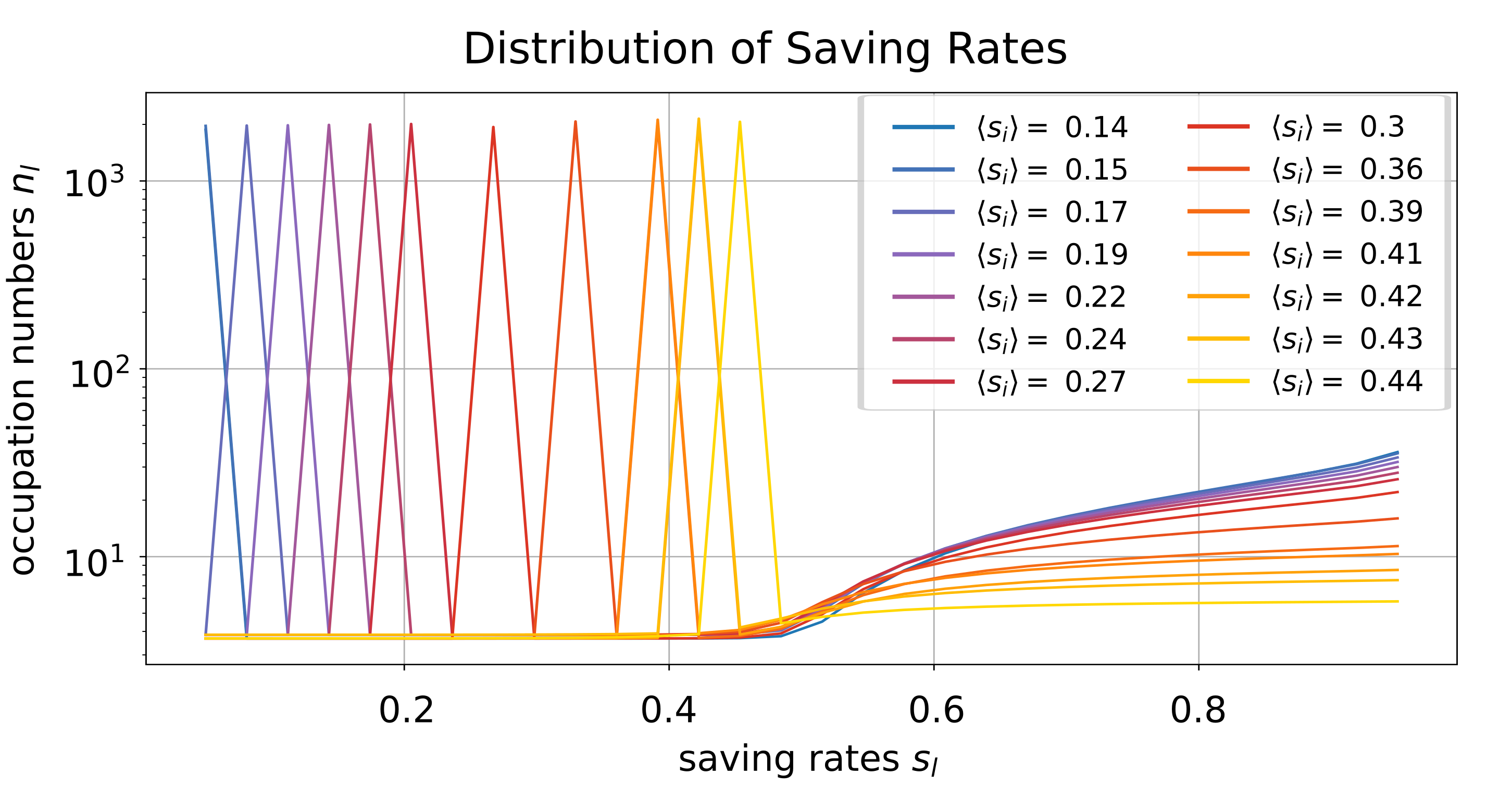}
    \caption{The saving rate distributions for all the fixed points in Fig.\ref{bifurcations} in the main text for $\beta=50$, labelled by their mean saving rate. Stable states correspond to (almost) each of the available levels below $s=0.5$.  When the dominant level has a higher saving rate, the tail of the distribution is less pronounced. Parameters as in Fig.\ref{bifurcations}(b) from the main text ($M=30$).}
    \label{many_levels}
\end{figure}

As shown in the main text, we find more fixed points, when considering the case with more than 5 available saving rate levels. In this case, we find that the new fixed points correspond to a similar situation as in the case $M=5$ mainly studied in the paper. Each of the fixed points corresponds to a situation where the majority of agents sit in a level with $s_l<0.5$ (Fig.\ref{many_levels} for $M=30$). Now that we have a higher resolution of the saving rates, we see that the second group of high savers is distributed along all the saving rate levels $s_l>0.5$. Notably a higher mean saving rate $\langle S_i \rangle$ corresponds to a smaller group of high savers and thus lower capital inequality as measured by the coefficient of variation, see below (Fig.\ref{produvtion_30_levels}).

\begin{figure}
    \centering
    \includegraphics[width=.9\textwidth]{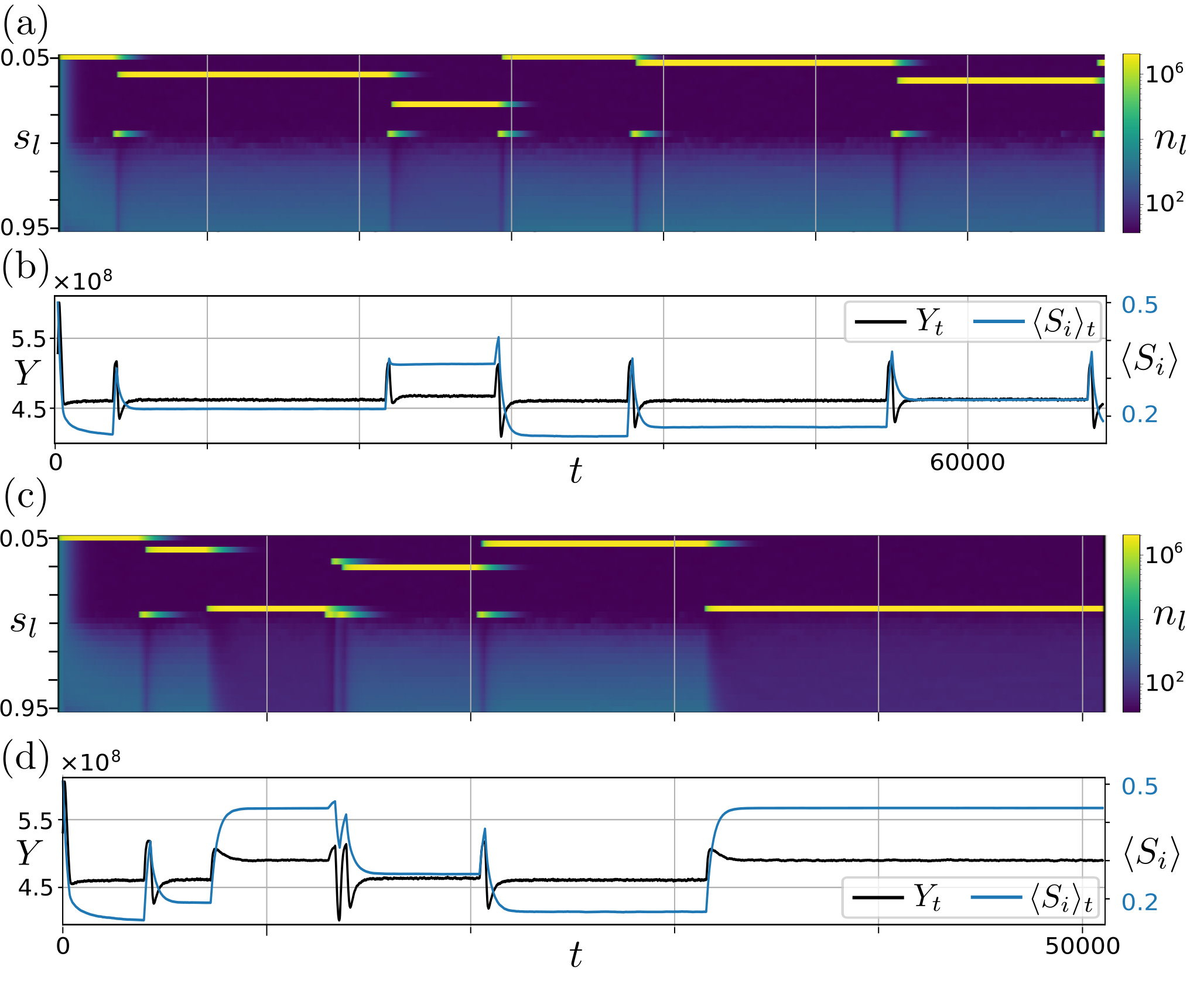}
    \caption{
    Two time series of the saving level occupation numbers (a), (c) and the resulting mean saving rate $\langle S_i \rangle$ and economic production $Y$ (b), (d) with $M = 30$ available saving rates. The two different realizations illustrate the more complex oscillatory behaviour to which the high degree of multistability gives rise to. We find metastable switching between the stable states in Fig.~\ref{many_levels} and short excursions to several unstable states. The realizations can remain at high saving rates near s = 0.5 for longer times and the transitions are accompanied by high fluctuations in economic production. All transitions start with a short
increase in production, but only some are followed by a recession. Parameters as in Fig.\ref{bifurcations}(b) from the main text, with $N=5.29 \times 10^6$.}
    \label{30_level Time Series}
\end{figure}

With the addition of noise, we again observe switching between the (now)  metastable states, however, in a much more complicated setting. Instead of switching between two states, we have 11 states and the switching dynamics becomes much more complex (Fig. \ref{30_level Time Series}). It is not clear which state the system will occupy after leaving a given state, and it would be interesting to study the oscillatory behavior of this more complicated model, and what the effects on economic growth would be.

\begin{figure}
    \centering
    \includegraphics{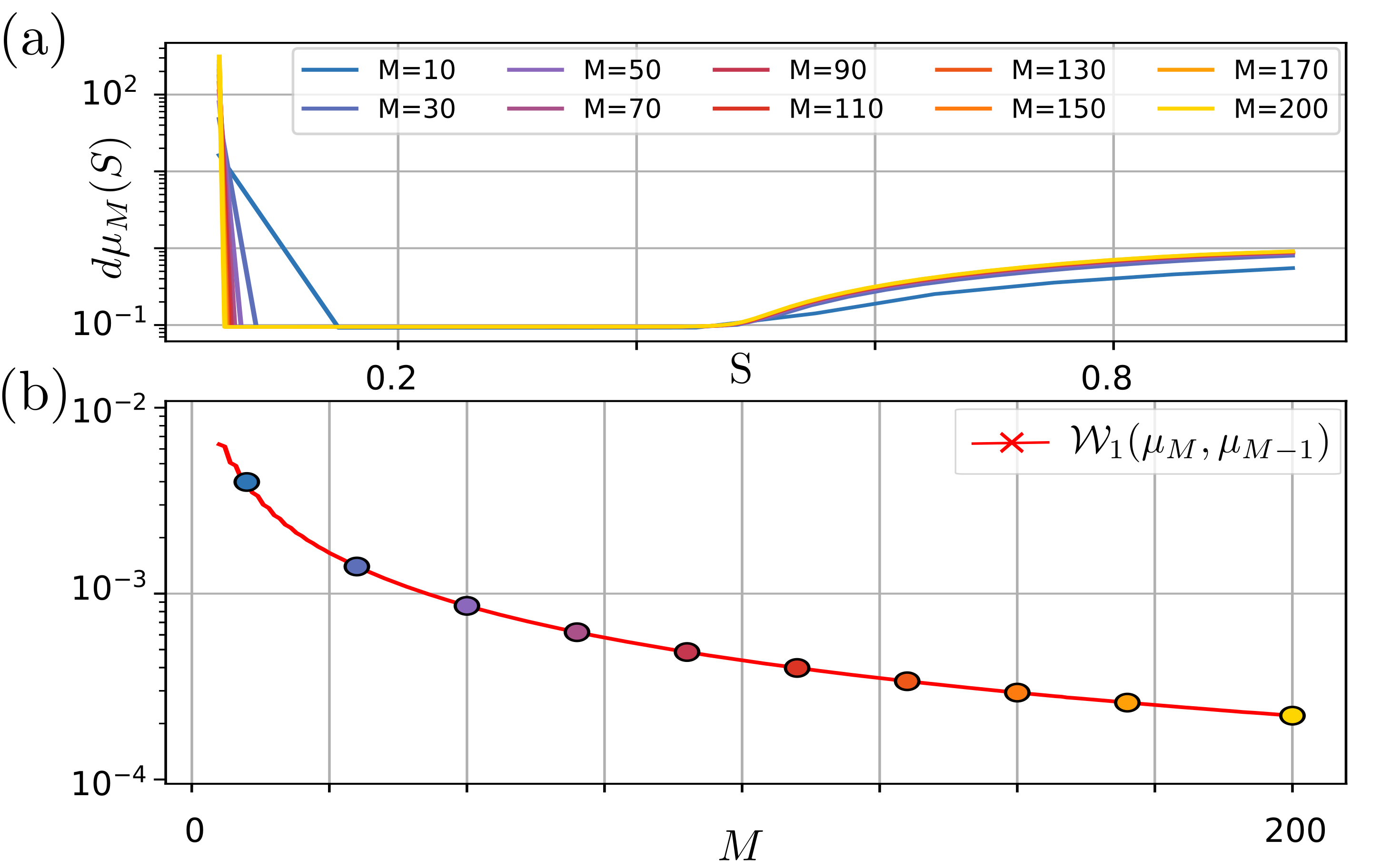}
    \caption{(a) The densities of the saving rate distributions $\mu_M(S)$ (plotted as continuous functions for visualization purposes) corresponding to the state with the smallest mean saving rate, for various values of $M$. (b) Wasserstein distance $\mathcal{W}_1$ between the saving rate distributions of each value of $M$ and the previous value $M-1$. 
    Each density in (a) corresponds to a specific $M$ in (b) indicated by the circles. As $M$ increases, there seems to be a limiting distribution. Parameters as in Fig \ref{many_levels}.}
    \label{iter distance}
\end{figure}

\begin{figure}
    \centering
    \includegraphics{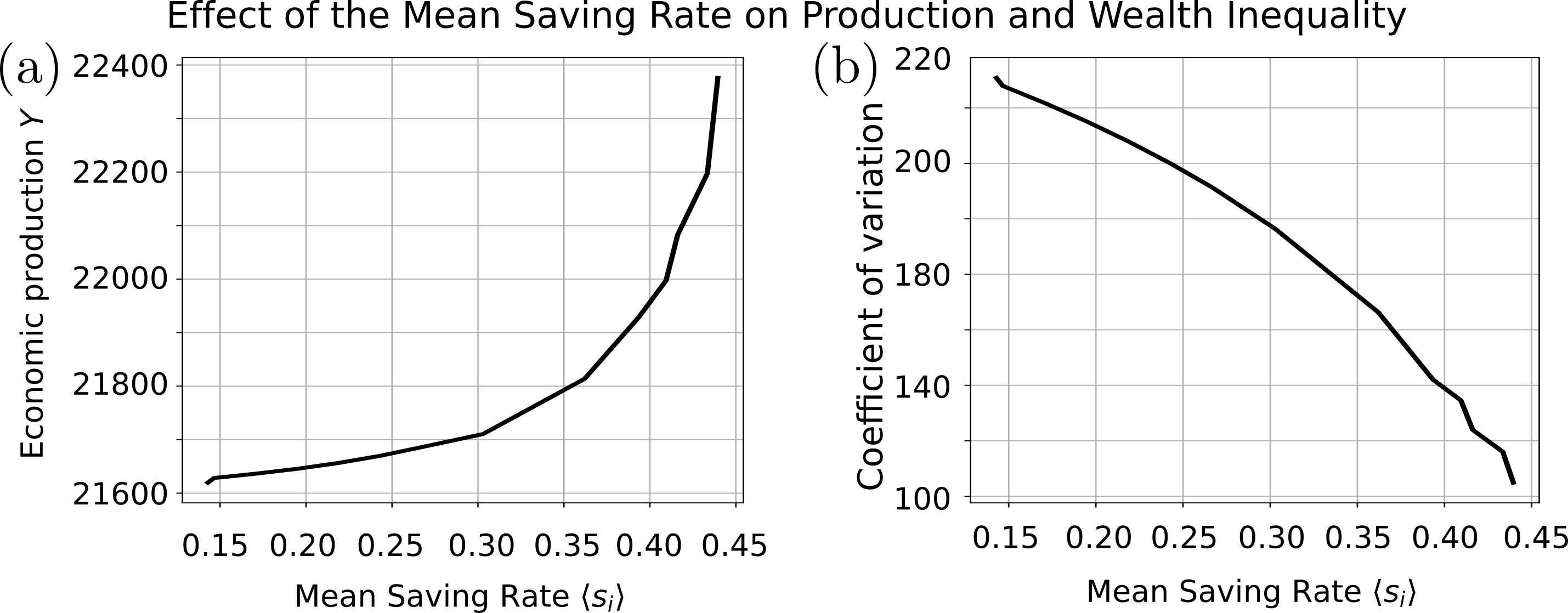}
    \caption{(a) Production and (b) coefficient of variation of the capital distribution for each stable state in Fig. \ref{many_levels} ($M=30$). The production increases slightly with larger mean saving rate, the coefficient of variation as a measure of capital inequality is reduced significantly.}
    \label{produvtion_30_levels}
\end{figure}

The state where a vast majority of agents occupy the lowest available saving rate is the most prominent state for a large variety of parameters, so we checked the existence of this state for up to $M=200$ (Fig.~\ref{iter distance}). The shape of the distributions does not change much for increasing $M$, and their general features are well captured by the case with lower $M$.

To get an estimate of the behavior for large $M$, we use an iterative process by (1) generating the fixed point with $M$, (2) then creating a new discretization with $M+1$ available saving rates, (3) using the interpolated result from the previous step as initial conditions and (4) integrating forward to obtain the new fixed point. This gives us a sequence of distributions $\lbrace \mu_M \rbrace_{M\in\mathbb{N}}$
\begin{align}\label{dist_M}
     \mu_M = \sum_{l=1}^M \frac{n_l}{N}\delta_{s_l}
\end{align}
where $\delta$ is the Dirac measure. The idea is that we would like this to be a fixed-point iteration.

To compare the resulting saving rate distributions, we make use of the Wasserstein distance 
\begin{align}
\mathcal{W}_1(\mu, \tilde{\mu}) = \underset{\pi \in \Gamma}{\inf}\int_{\left[0,1\right]^2} \vert S-\tilde{S}\vert d\,\pi(S, \tilde{S})
\end{align}
where $\Gamma$ is the set of probability measures that have $\mu, \tilde{\mu}$ as their marginal. For the one-dimensional case \cite{Wasserstein}, the Wasserstein distance can be written as
\begin{align}
    \mathcal{W}_1(\mu, \tilde{\mu}) =\int_{\left[0,1\right]} \vert F_{\mu}(S)-F_{\tilde{\mu}}(S)\vert
 \,dS \end{align}
in terms of the respective cumulative distributions $F_{\mu}$ and $F_{\Tilde{\mu}}$, if $\mu, \tilde{\mu}$ have finite moments. This formulation is particularly useful because it allows us to compare the distributions (\ref{dist_M}) for different $M$, irrespectively of the saving rate levels taking different values.

 To investigate the behaviour of the sequence $\lbrace \mu_M \rbrace_{M\in \mathbb{N}}$, given by (\ref{dist_M}) as $M\rightarrow\infty$ In Fig.~\ref{iter distance}b we plot the difference  $\mathcal{W}_1(\mu_{M-1},\mu_M)$ between iterations. We can see that the difference between iterations becomes small very fast.

If we assume that this is a convergent fixed point iteration, it seems that the distribution for $M=200$ may already be considered as representative of the limit. 

So far we have only used equidistant saving rates between $0.05$ and $0.95$, but we have also verified that the choice of saving rate levels has no impact on this state by setting $M=100$ and choosing arbitrarily $s_l \in [0.001, 0.999]$.

 If we look at the economic production of each stable state in Fig.~\ref{many_levels}, we can see that the new stable states resulting from a larger $M$ induce more intermediate levels of production, which gives a better approximation of the production $Y$ as a function of mean saving rate $\langle S_i \rangle$. This indicates that the coarse-graining of the saving rates results in a satisfactory coarse-graining of the production, which captures the lowest and highest production levels already for $M=5$.

 An advantage of the discretization of saving rates is also that only course-grained information is needed on the saving rates. Furthermore in our macroscopic model, each agent only needs to know the statistical distribution of consumption and saving rate in the global population, while in the agent-based model  ([5] in the main text) rather precise knowledge of the individual consumption and saving rates is required.
\section{Spikes of Mean Consumption}
In order to understand how the sudden consumption  spikes arise, we need to discuss the different mechanisms that can induce an increase of mean consumption. The first possibility is an increase in mean capital due to the market dynamics, which is clearly not the case here, since the returns and thus also the wages are almost constant during the spike in Fig. \ref{few_agents_cause transitions}(c) in the main text. The only other way the mean consumption can increase is through the influx of capital that other agents carry to a given level. 
For a given time series we can directly calculate the change of capital $\Delta_k \langle K_i\rangle_l$ due to these mechanisms
\begin{align}
    \Delta_{k}m_l^1 = \int_{t_0}^{t_1} dt'\, \frac{m^1_k - m^1_l}{n_l}\alpha_{kl} \quad \text{for }k\neq l\\
    \Delta_{l}m_l^1 = \int_{t_0}^{t_1} dt'\, (rs_l-\kappa)m_l^1 \quad \text{for } k = l
\end{align}
Where $\Delta_l m_l^1$ denotes the changes due to the market dynamics. Note that the integrants sum up to give the right-hand side of Eqn.~(\ref{SDE-non-central-moments}) in the main text, for $p=1$.
Integrating the individual contributions to the change in mean capital over the highlighted time period of the spike in Fig. \ref{few_agents_cause transitions}(c) in the main text shows that during the spike the greatest contribution comes from agents that switch from $s_5$ to $s_2$, and all the other contributions are either negligible or negative in the case of agents coming from $s_1$, where the agents carry less capital on average (Fig. \ref{few_agents_cause transitions}(e) in the main text).

In the same way we can split the terms in $d\langle C_i\rangle_l/dt$ and verify that the transfer of capital from the highest saving rate level is actually responsible for the increase in mean consumption during the spike (Fig. \ref{few_agents_cause transitions} e, f in the main text). Similarly, we can calculate the ratio of agents that are involved in the capital transfer from $s_5$ to $s_2$ during the spike in Fig.\ref{few_agents_cause transitions} (in the main text) 
\begin{align}
    \frac{1}{N}\int_{t_0}^{t_1}dt'\,\alpha_{52} = 0.14 \%
\end{align}

\section{Correlation Time}
For a Markovian stochastic process $\mathbf{X}_t$ the auto-correlation 
\begin{align}
    C(\tau) &= \mathbb{E}[\tilde{\mathbf{X}}_t\tilde{\mathbf{X}}_{t+\tau}] \quad \text{where: } \tilde{\mathbf{X}}_t = \mathbf{X}_t - \mathbb{E}[\mathbf{X}_t]
\end{align}
measures the correlation of a process with a shifted version of itself. So we can immediately see that this would be useful to detect possible periodicities in the process. If we have high local maxima in the auto-correlation for a set of shifts $\tau = nT$ for some $T\in \mathbb{R}, n\in \mathbb{N}$ this would indicate that realizations of the process are likely to be $T$-periodic. The simplest is just a sine wave with added noise. Even if the noise is quite strong, and the oscillations might be hard to identify by just looking at the time series, the auto-correlation will often show the periodicity.

However when the oscillations do not follow a simple periodicity, spotting these becomes much harder, as the auto-correlation for a Markov process also decays exponentially and eventually the $\mathbf{X}_t$ and $\mathbf{X}_{t+\tau}$ become uncorrelated.

An improved measure for coherence is the correlation time $\tau_{corr}$ of a signal. However also here there are two different definitions in the literature \cite{Geffert2014_SM, Pikovsky_SM}. Here we choose
\begin{align}
    \tau_{corr} = \frac{1}{C(0)}\int_0^\infty \vert C(\tau)\vert \,d\tau
\end{align}
where we have used the physical definition of the autocorrelation function \cite{Geffert2014_SM}. To see how the correlation time measures the coherence of a process,  we can make use of the relation of the spectral power density
\begin{align}
    S(\omega) = \mathbb{E}[ \vert\mathcal{F}[\mathbf{X}_t](\omega)\vert^2]
\end{align}
and the auto-correlation function, given by the Wiener–Khinchin theorem
\begin{align}
    S(\omega) = \mathcal{F}[C(\tau)](\omega)
\end{align}
Note that the Fourier transform of neither $C$ nor $\mathbf{X}_t$ in general have to exist and the theorem can be stated more broadly. But for most physical applications we can assume their existence.

\begin{figure}
    \centering
    \includegraphics[width=.7\textwidth]{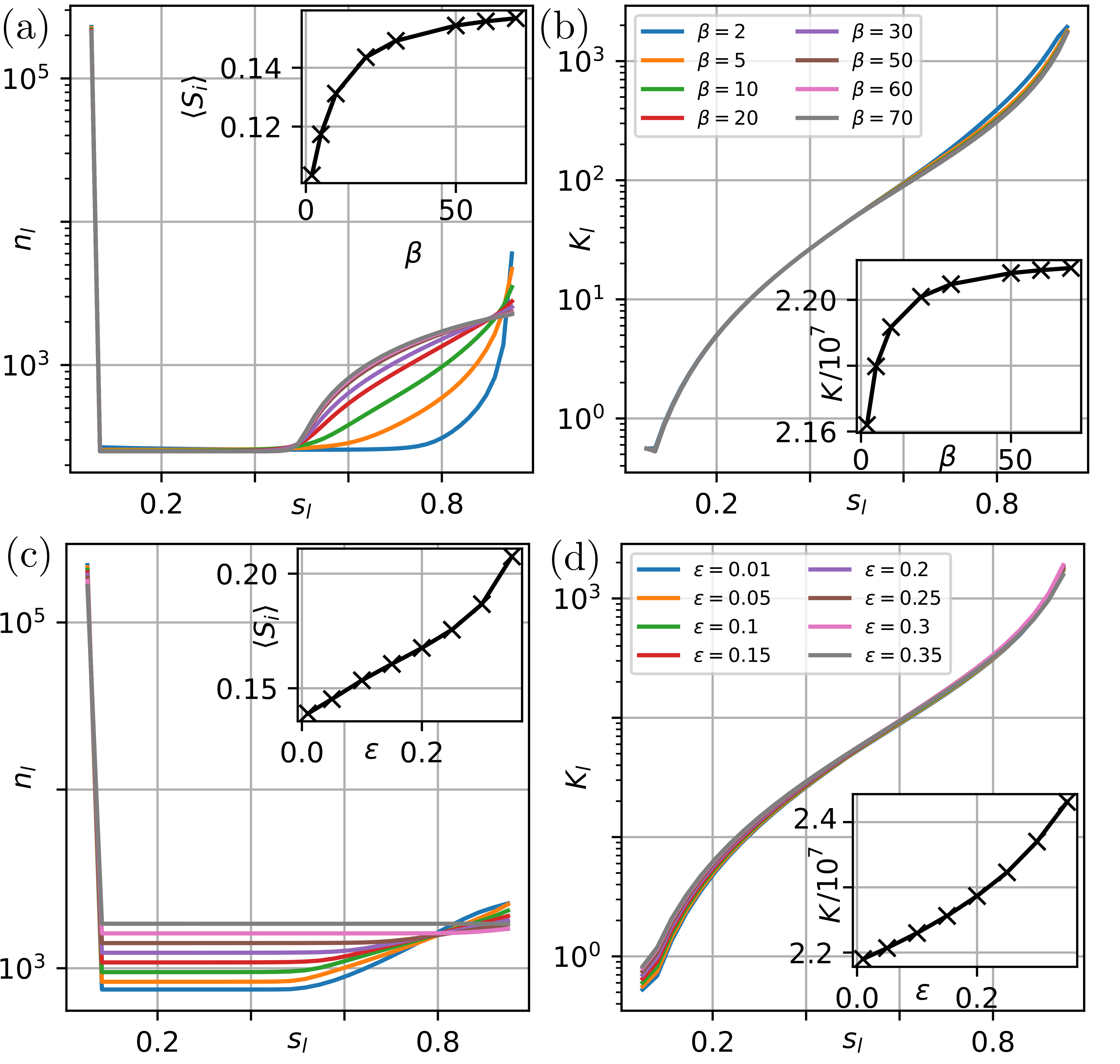}
    \caption{(a),(c) Saving rate distribution and (b),(d) capital distribution of the stable state where most agents adopt the lowest available saving rate for $M=50$. In panels (a),(b) $\epsilon=0.05$ and various values of $\beta$; in (c),(d) $\beta=30$ and various values of $\epsilon$. The general shape is not affected by larger choices of $M$. The insets show the mean saving rate $\langle S_i\rangle$ and the aggregate capital $K$ as a function of $\beta$, or $\epsilon$, respectively. Other parameters as in Fig.~\ref{bifurcations} in the main text.}
    \label{parameter_variation_beta}
\end{figure}

\section{Parameter Variations}
Here we vary the parameters related to wealth distribution and agent
decision-making. In Fig. \ref{parameter_variation_beta}(a),(b) we plot the saving rate and capital distribution for $M=50$ and several values of the inverse temperature $\beta$ again for the most prominent state of the model. Counter-intuitively, the saving rate distribution becomes much more spread out at higher saving rates, when $\beta$ is increased. For  $\beta=2$ the distribution drops faster than exponentially when moving away from the highest saving rate. As $\beta$ increases, the drop-off becomes less and less pronounced, until it is almost linear for $\beta=50$.
The reason for this is that, as $\beta$ increases, more weight is put on imitating the highest saving rate, instead of following the majority, which leads to the increase in mean saving rate, which in turn increases the economic output, and reduces the return rate. This diminishes the advantage of maintaining a saving rate close to $s=0.95$ and allows the agents with slightly lower saving rate to compete with the "high savers".

Notably the distribution of aggregate capital $K_l= \sum_{\lbrace i:S_i=s_l\rbrace} K_i = n_l m_l^1$ is very resilient against changes in $\beta$. The effect of increased occupation numbers is cancelled out by the diminishing returns due to the increase of aggregate capital $K$, which increases by roughly 2\% over the whole range of $\beta$.

In Fig. \ref{parameter_variation_beta}(c),(d) we plot the saving rate and capital distribution for $M=50$ and several values of $\epsilon$ again for the most prominent state of the model. As the exploration tendency $\epsilon$ of the agents increases, the saving rate distributions approach a uniform distribution and the peaks become less pronounced. There is almost no influence of $\epsilon$ on the capital distribution.
Again we observe a strong correlation between mean saving rate $\langle S_i\rangle$ and aggregate capital $K$. 

Note that the lowest value of $\epsilon=0.01$ can be used in the deterministic limit, but with the addition of noise, we run into numerical issues, since many of the terms in the model scale with $1/n_l$, and therefore we need to employ a slightly larger value of $\epsilon$. We made sure that with such small $\epsilon$ very similar behaviour is  recovered as in the agent-based model ([5] in the main text) where exploration of this form is not present. We believe that the main effects in our model are due to imitation behaviour and not due to exploratory behaviour.

\section{Bibliographical Notes on Consumption Behavior}
Some empirical studies exist on the interplay of economic inequality and conspicuous consumption behavior that qualitatively relate to assumptions and observations made in our model.

Previous empirical research has identified that people or households are not only concerned with absolute consumption. There is considerable evidence that relative consumption (compared to their peers) plays an important role in creating status-seeking behaviour and status anxiety \cite{CHARLES201314, ALPIZAR2005405, carllson2007, SIVANATHAN2010564}, which often expresses itself through investing in resources for future returns that increase status (consumption in this case) \cite{Velandia-Morales}. This fact illustrates the importance of the social network between households, and that their interactions can drive saving decisions.
In the literature, status anxiety has mostly been studied as a consequence of inequality, but \cite{Velandia-Morales} points out that status anxiety can also trigger consumption decisions and the directions of this relationship are not clear. Although households can invest in anticipation of future consumption increases, status anxiety caused by other households' consumption can lead to a decision to increase spending, even though one could argue that households should also be able to anticipate the rapid decline of capital stock.  

In \cite{Household_Saving_Class_Identity} evidence is presented that an alleged high degree of mobility (resulting in status-seeking behaviour and conspicuous consumption) is connected to an overall lower-than-expected saving rate in the USA. And \cite{CHARLES201314} finds that 'Results support depictions of expenditure cascades, where spending by those better off ratchets up local standards of 'normal' and socially acceptable living.' This is qualitatively very similar to our result that the states with low saving rates are very prominent and that their dominance is driven by a constant flow of agents from the wealthy high saver group driving up the mean consumption.

In \cite{CHARLES201314} it is also hypothesized that  "growing economic inequality and positional consumption may be a self-reinforcing process, which is a feedback loop that emerges in our model if we argue that our model contains positional consumption in the sense that one's own consumption is only valued by comparison with other agents and not by the absolute level of consumption.

In our model the agent imitates someone else's saving rate in the expectation that this savings strategy will lead to a similar consumption level {\em eventually} because of the (basically correct) anticipation of the capital growth effect of the savings rate. So the model is not completely unrealistic in that it assumes some degree of anticipation, farsightedness, and patience on the side of the agents. On average the agents allow a time $\tau$ for a new strategy to prove effective. This farsightedness is much less prominent than in classical growth models, where agents are assumed to perform a long term optimization, taking into account (discounted) payoffs arbitrarily far into the future.

The agents switching during the first spike induce capital flow between the levels, which can be compared to the change in capital/consumption due to market dynamics.
This comparison yields that the only significant contribution to the sudden increase in consumption can be associated to agents switching from the wealthy high saver group to a lower saving rate level. The subset of the $0.1 \%$ of agents is the subset of agents taking part in this particular transition during the initial increase in consumption, which we deem responsible for the drastic effects on macroeconomic variables.

Our model provides insight into the key nonlinear mechanisms that drive tipping between different macroeconomic states, which is an important task when evaluating the effect of policies. Since we show that there is a strong observable precursor of the metastable switching this opens up the possibility for a control scheme that is activated once the initial consumption spike is observed and keeps the system from tipping into an undesirable state. Since a higher average saving rate coincides with increased economic production and reduced capital inequality,  this analysis might help in designing such policies.

\end{document}